# New avenues for characterizing individual mineralized collagen fibrils with transmission electron microscopy

Tatiana Kochetkova,[a*] Stephanie M. Ribet,[b] Lilian M. Vogl,[bde] Daniele Casari,[c] Rohan Dhall,[b] Philippe K. Zysset,[a] Andrew M. Minor,[bd] Peter Schweizer[be*]

[a] ARTORG Center for Biomedical Engineering Research, University of Bern, Bern CH-3010, Switzerland;

[b] National Center for Electron Microscopy (NCEM), The Molecular Foundry, Lawrence Berkeley National Laboratory, Berkeley, CA 94720, USA;

[c] Laboratory for Mechanics of Materials & Nanostructures, Empa - Swiss Federal Laboratories for Materials Science and Technology, Thun CH-3603, Switzerland;

[d] Department of Materials Science and Engineering, University of California Berkeley, CA 94720, USA;

[e] Max Planck Institute for Sustainable Materials, Düsseldorf 40237, Germany

* Corresponding Authors: tatiana.kochetkova@unibe.ch, p.schweizer@mpi-susmat.de

## Abstract

Bone serves as a remarkable example of nature's architectured material with its unique blend of strength and toughness, all at a lightweight design. Given the hierarchical nature of these materials, it is essential to understand the governing mechanisms and organization of its constituents across length scales for bio-inspired structural design. Despite recent advances in transmission electron miscoscopy (TEM) that have allowed us to witness the fascinating arrangement of bone at micro-down to the nano-scale, we are still missing the details about the structural organization and mechanical properties of the main building blocks of bone – mineralized collagen fibrils (MCFs). Here, we propose a novel approach for extracting individual MCFs from nature's model material via a dropcasting procedure. By isolating the MCFs onto TEM-compatible substrates, we visualized the arrangement of organic and mineral phases within the individual MCFs at the nanoscale. Using a 4D-STEM approach, the orientation of individual mineral crystals within the MCFs was examined. Furthermore, we conducted first-of-its-kind *in situ* tensile experiments, revealing exceptional tensile strains of at least 8%, demonstrating the intricate relationship between structural organization and the mechanical behavior of MCFs. The capabilities of TEM allow us to resolve MCF organization and composition down to the nanoscale level. This new knowledge of the ultrastructure of the bone-building blocks and the proposed sample extraction and *in situ* mechanical testing opens up new avenues for research into nature's inspired material design.

## Introduction

Bone is a complex natural material exhibiting an exceptional combination of strength and toughness while being lightweight. These remarkable material properties are triggered by the evolutionary adaptation of bone to provide mechanical support and organ protection in the vertebrates.[1] While bone is composed of ductile organic and brittle inorganic phases, the hierarchical arrangement of the constituents from macro- down to the nanoscale gives rise to the outstanding mechanical properties (**_Figure 1_**).

Mineralized collagen fibrils (MCFs) are the building blocks of bone tissue, reaching about 30-300 nm in diameter and up to hundreds of microns in length. MCFs are formed from staggered arrays of tropocollagen molecules intertwined with hydroxyapatite nanocrystals, stabilized by hydrogen bonds and cross-links.[2,3] Crystal plates of hydroxyapatite (HA) assemble in the gap between the molecules, reaching tens of nanometers in length and 1-2 nm in thickness. This staggered arrangement creates an observable periodicity within the MCF known as the D-band with the reported repetition step of 67 nm.[4,5]

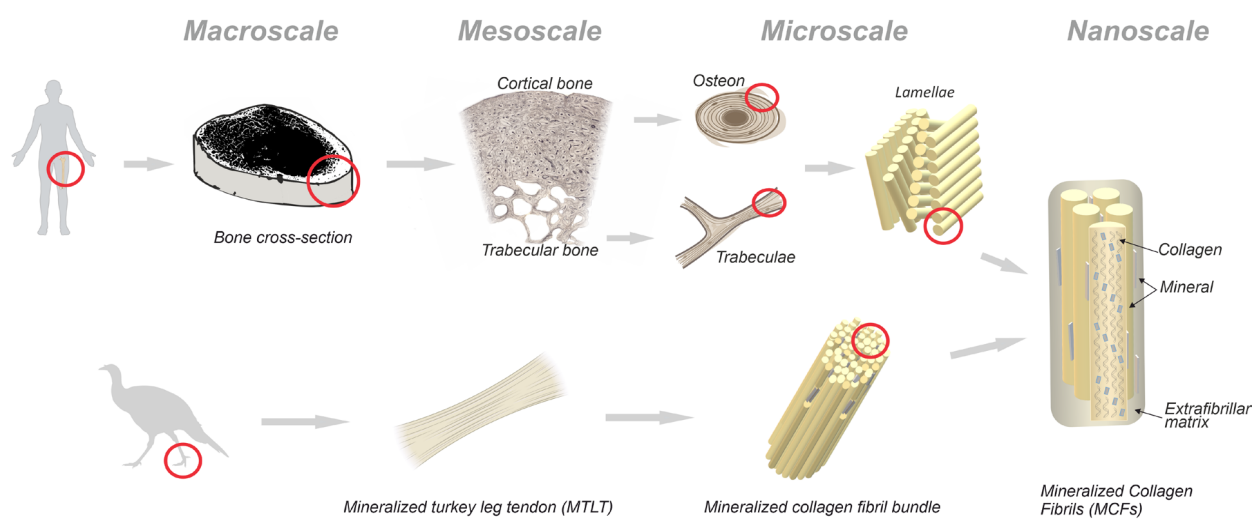

*Figure 1. Human lamellar bone exhibits complex hierarchical organization spanning from macro- to nanoscale levels. Mineralized collagen fibrils (MCFs) are considered the building blocks of bone, made of staggered collagen tropomolecules with intra- and extramolecular inclusions of the mineral agglomerates. On the other hand, the mineralized tendon from the turkey leg is made of a uniaxial bundle of MCFs, making it an attractive model to study the building blocks of bone – the MCFs.*

Direct observation of mineralized collagen fibrils (MCFs) in lamellar bone requires nanometer-scale resolution, strong collagen–mineral contrast, and, ideally, preserving the native hydrated structure. Recent studies have employed high-resolution TEM and STEM for mineral–collagen imaging,[6,7] electron tomography for nanoscale 3D reconstructions,[8,9] FIB–SEM for volumetric mapping of mineral organization,[10,11] and synchrotron-based conventional SAXS/WAXS or even their tensor tomography modalities for orientation and density mapping.[12–14] Across these modalities, common limitations stem from sample preparation (dehydration, embedding, staining, etching), which can alter native fibril morphology; trade-offs between resolution and field of view; and restricted imaging volumes in high-resolution 3D methods. As a result, while the hierarchical organization of MCFs *in situ* is well documented, direct visualization of fully isolated, intact MCFs in their native state remains scarce, leaving key aspects of individual fibril geometry and mineralization unresolved.

Other studies have attempted to synthesize MCF-like structures by mineralizing non-mineralized collagen fibrils *in vitro* using biomimetic approaches such as polymer-induced liquid-precursor mineralization, simulated body fluid immersion, or controlled calcium phosphate precipitation.[15–17] While these synthetic constructs often reproduce the expected nanoscale organization of apatite platelets within the collagen gap zones, they inevitably lack the natural developmental and compositional context of bone. Consequently, their structure and properties, though informative, cannot fully substitute for direct observations of native MCFs.

A new methodological approach is therefore required — one that enables *in situ*, high-resolution visualization of natural MCFs in their unaltered state, preserving both mineral and collagen phases without destructive sample preparation. Such an approach would close the critical knowledge gap between synthetic models and the true nanoscale architecture of bone.

Here, we propose the mineralized turkey leg tendon (MTLT) as a source of MCFs, which are more straightforward to extract and easily accessible. The chemical composition as well as the crystal organization in MTLT closely approximate those found in bone.[18–23] But the primary advantage of MTLT tissue is its simplified fiber arrangement, which differs from the complex structure of mammalian cortical bone. MTLT is composed of densely packed collagen fibrils, which are strongly aligned with the tendon axis, meaning that global and local fibril orientations are highly correlated.[24,25] Following the structural and compositional similarities, MTLT has also been used as a valuable model to study the mechanical properties of bone at the microscale.[25–28] However, whether this can be extended to nanoscale studies on individual MCFs remains to be investigated.

This work ambitiously pursues two goals: (i) TEM imaging of isolated MCFs and (ii) *in situ* TEM observations of the MCFs deformation. For this, we develop a novel protocol for the MCFs extraction and image them via the high-resolution TEM modalities. Ultimately, the isolated MCFs imaging allows the *in situ* mechanical tests, elucidating the fundamental processes of fibril deformation, and has the potential to pave the way for creating innovative bio-inspired nanostructured materials.

# Results

## Imaging the individual MCF

The dropcasting protocol allowed the extraction of the MCFs from the bulk MTLT sample, preserving their native arrangement. For this, individual MCFs were extracted from the bulk MTLT samples via mechanical splitting and ultrasonification. The supernatant with the intact MCFs was then transferred to the TEM grids via dropcasting (***Figure 2***). The two types of grids were used: (i) a copper TEM grid with lacey carbon films for static imaging or (ii) a copper tensile stripe with support film in the central window.

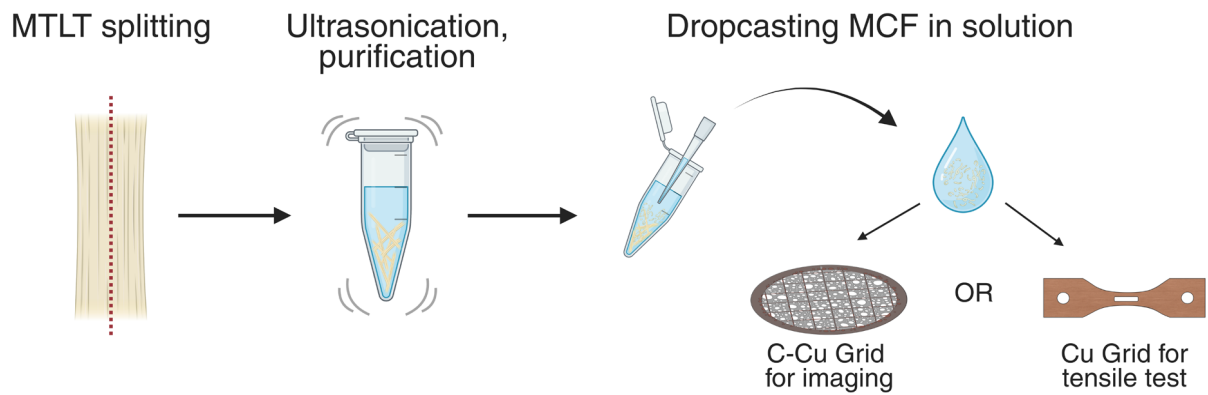

*Figure 2. Hydrated mineralized turkey leg tendons (MTLT) were split along the main axis and ultrasonicated; the resulting solution with mineralized collaged fibrils (MCFs) is dropcasted on (i) a copper TEM grid with lacey carbon films for imaging or (ii) a copper tensile stripe with support film in the central window. Figure created using BioRender.com.*

## Origins of nanoscale periodicity

Individual MCFs were observed on the lacy carbon support film of the TEM grids, with their characteristic compositional contrast clearly resolved in both high-angle annular dark-field (HAADF) STEM images and corresponding energy-dispersive X-ray spectroscopy (EDX) elemental maps (***Figure 3***). The periodic arrangement of mineral phases was evident in both imaging modes, yielding an average D-period of $68.64 \pm 0.16$ nm, as determined from the elemental intensity profiles of five MCFs. This value is slightly larger than the canonical 67 nm reported for native collagen.[4] MCFs exhibited varying degrees of mineralization, with calcium-to-phosphorus (Ca/P) ratios ranging from 0.26 to 1.63; the upper value being close to the theoretical Ca/P ratio of 1.67 for stoichiometric hydroxyapatite.[29] Analysis of a limited dataset (n=5) indicated a negative correlation between the D-period and Ca/P ratio ($R^2 = 0.78$, ***Figure S 2***), suggesting that higher mineral content is associated with a slight reduction in fibril's D-period.

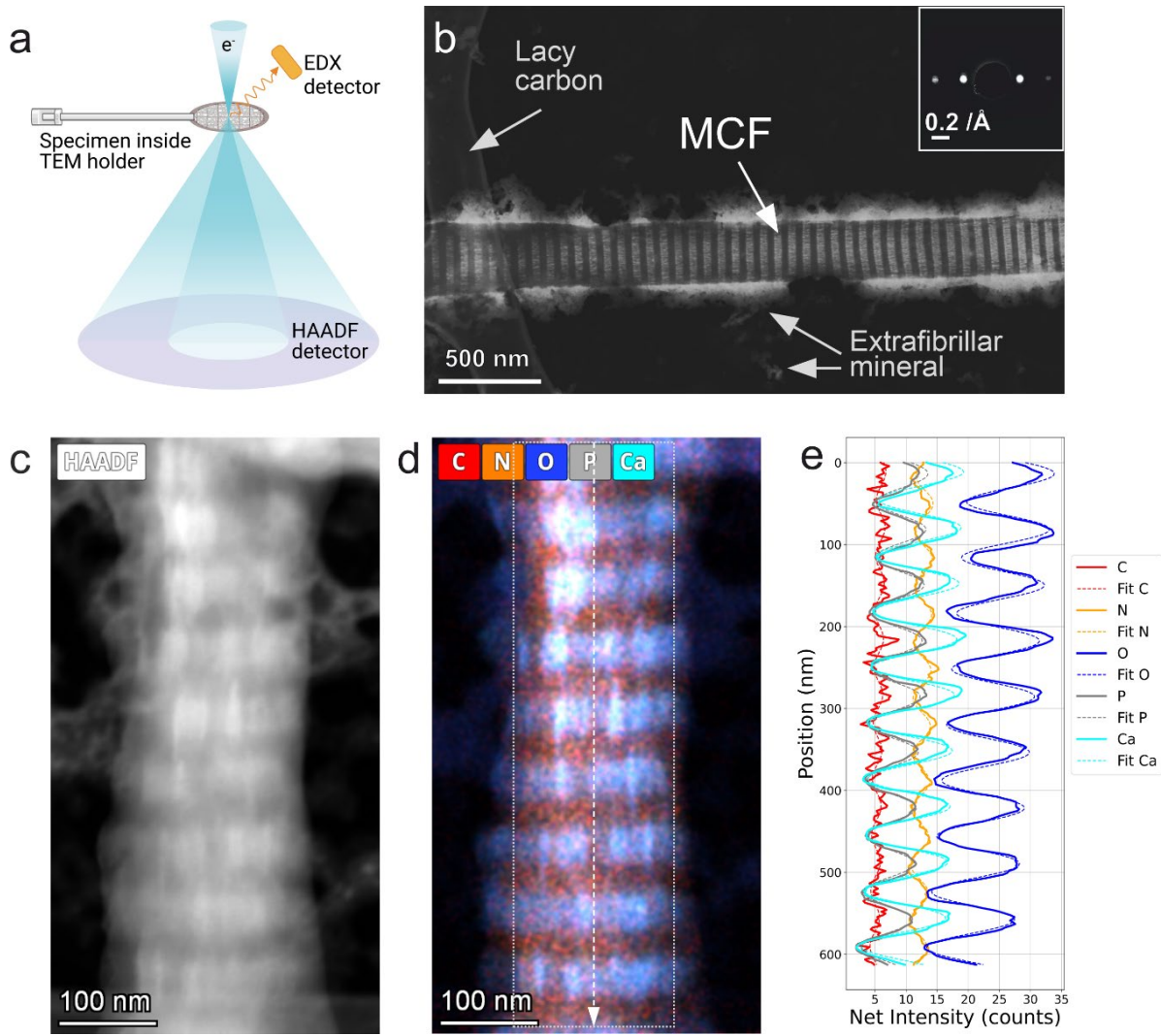

*Figure 3. MCF appearance in TEM. a) Schematic of TEM imaging with the energy-dispersive X-ray spectroscopy (EDX) detector and the high-angle annular dark-field (HAADF) detector. Overview HAADF STEM image (b) demonstrates visible periodicity from the electron absorbance, while the electron diffraction highlights the preferred crystallographic orientation of the HA along the main fiber axis. Close-up HAADF (c) captured the periodicity of the MCF organization along the main axis, followed by periodic compositional distribution as visible from EDX maps (d) and their fits (e).*

## Crystall localization and orientation

Four-dimensional scanning transmission electron microscopy (4D-STEM) was used to probe the nanoscale structure of the mineralized collagen fibrils. In this method, a converged electron nano-beam is raster-scanned across the specimen, and a pixelated detector records a diffraction pattern at each probe position, enabling the reconstruction of spatially resolved structural maps even from the beam-sensitive materials like MCFs.[30] The 4D-STEM datasets obtained from individual fibrils revealed distinct diffraction features corresponding to crystalline mineral domains embedded within the collagen matrix.

To assess mineral orientation, we applied flow line mapping relative to the crystallite c-axis (the (002) reflection). The detailed data processing and reconstruction pipeline is summarized in ***Figure S 3***.In this approach, the angular position of the (002) diffraction spot is measured at each scan pixel, and the resulting orientation vectors are plotted as continuous lines across the real-space map, tracing the in-plane orientation of the crystallites (**Figure 4**). In six out of eight 4D-STEM scans, the flow line maps demonstrated moderate alignment of the hydroxyapatite c-axis to the MCF backbone, with an angular deviation of 2 ± 12° (Scans 1-6, **Figure S 3**). The remaining two scans revealed a broad range of hydroxyapatite orientations, suggesting the strong influence of the extrafibrillar mineral (Scans 7 and 8, **Figure S 3**). Moreover, variations in diffraction peak intensity across the scans indicated heterogeneity in local mineral content, in agreement with the variations in Ca/P ratio measured by EDX.

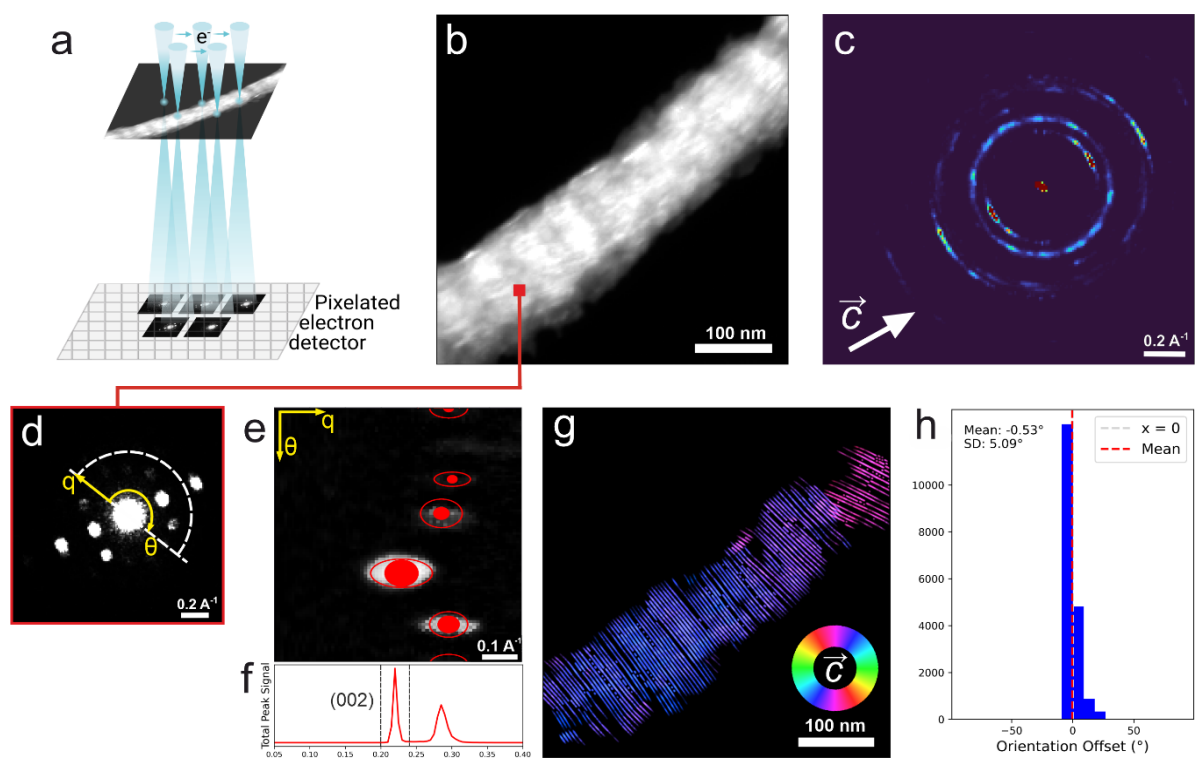

*Figure 4. 4D-STEM scan of a representative MCF. a) Schematic of 4D-STEM imaging principles. (b) Reconstructed MCF fibril image in the dark field. (c) Mean electron diffraction from the MCF fibril with a marked c-axis of hydroxyapatite. (d) Local electron diffraction from a single pixel marked in the dark field image (b). (e) Following the polar transformation as marked in (d), diffraction peaks are fitted, and the (002) reflection is selected for the image reconstruction (f). (g) The resulting flow line map depicts the in-plane orientation of the c-axis of the hydroxyapatite crystallite. (h) Histogram of the orientation offset of the hydroxyapatite crystallite relative to the fibril orientation.*

## Structurally guided mechanical properties

*In situ* tensile experiments on mineralized collagen fibrils (MCFs) were performed in a double-aberration-corrected FEI Titan microscope (TEAM I) using a single-tilt straining holder equipped with a custom-designed copper tensile stripe. Dropcasted MCFs were distributed across a nitro-cellulose support film in the central window of the tensile stripe. Quasistatic tensile loading was manually applied via the straining sample holder, inducing guided stretching of the support film and the attached MCFs.

We tracked the deformation progression of an individual MCF (**Figure 5**). This fibril exhibited crack initiation and propagation, revealing key deformation mechanisms under tensile stress. Crack nucleation occurred at the interface between the mineral-rich (gap zone, appears bright) and collagen-rich (overlap zone, appears dark) zones. Although nucleation took place outside the initial field of view, subsequent crack growth was recorded. The crack propagated along the overlap region (dark, collagen-rich), following a distorted path within this narrow collagen-rich zone.

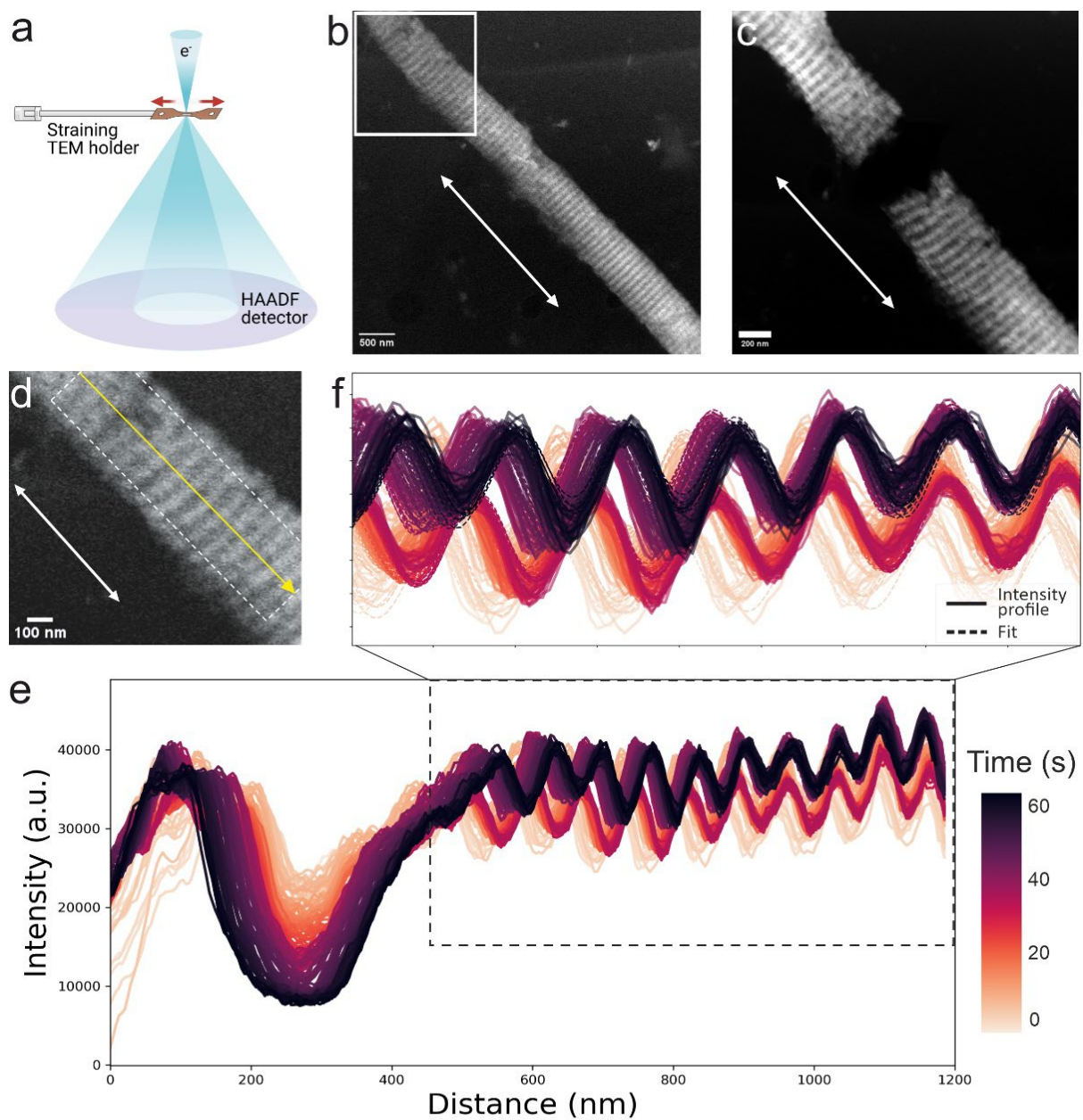

*Figure 5. In situ tensile testing of individual MCF. (a) Schematic of tensile testing in TEM with the high-angle annular dark-field (HAADF) detector. (b) Dark field image of the MCF fibril during the tensile test, with the highlighted region (d) of the crack initiation. (c) Dark field image of the fibril after the separation following the tensile test. (d) Zoomed-in image of the deformation region with the marked area of intensity measurements. (e) Resulting intensity profile across the MCF region highlighted in (c). (f) Zoomed in view on the intensity profile that was fitted for the D-period estimations during the tensile stretching.*

D-period tracking during crack progression provided further insight into the internal structural response of the fibril (**Figure 6**). Prior to the applied deformation (zero-strain), the D-period of the fibril near the crack opening was 69.5 ± 0.6 nm, as estimated from a series of 30 static dark-field images. Thus, the fibril was in a pre-strained state before we induced the stretching. Strain relaxation was observed as the crack opened, and this relaxation continued even after complete fibril separation at 40 seconds, ultimately reaching the D-period of 66.6 nm. While initial crack nucleation occurred outside the imaging area, the collected data during crack progression indicated that the MCF reached a remarkable strain of 8.2%, which may have been even higher at the moment of crack nucleation. Moreover, post-failure imaging of the fracture surface revealed an irregular morphology (**Figure 5 c**), suggesting active crack deflection at the organo-mineral interface.

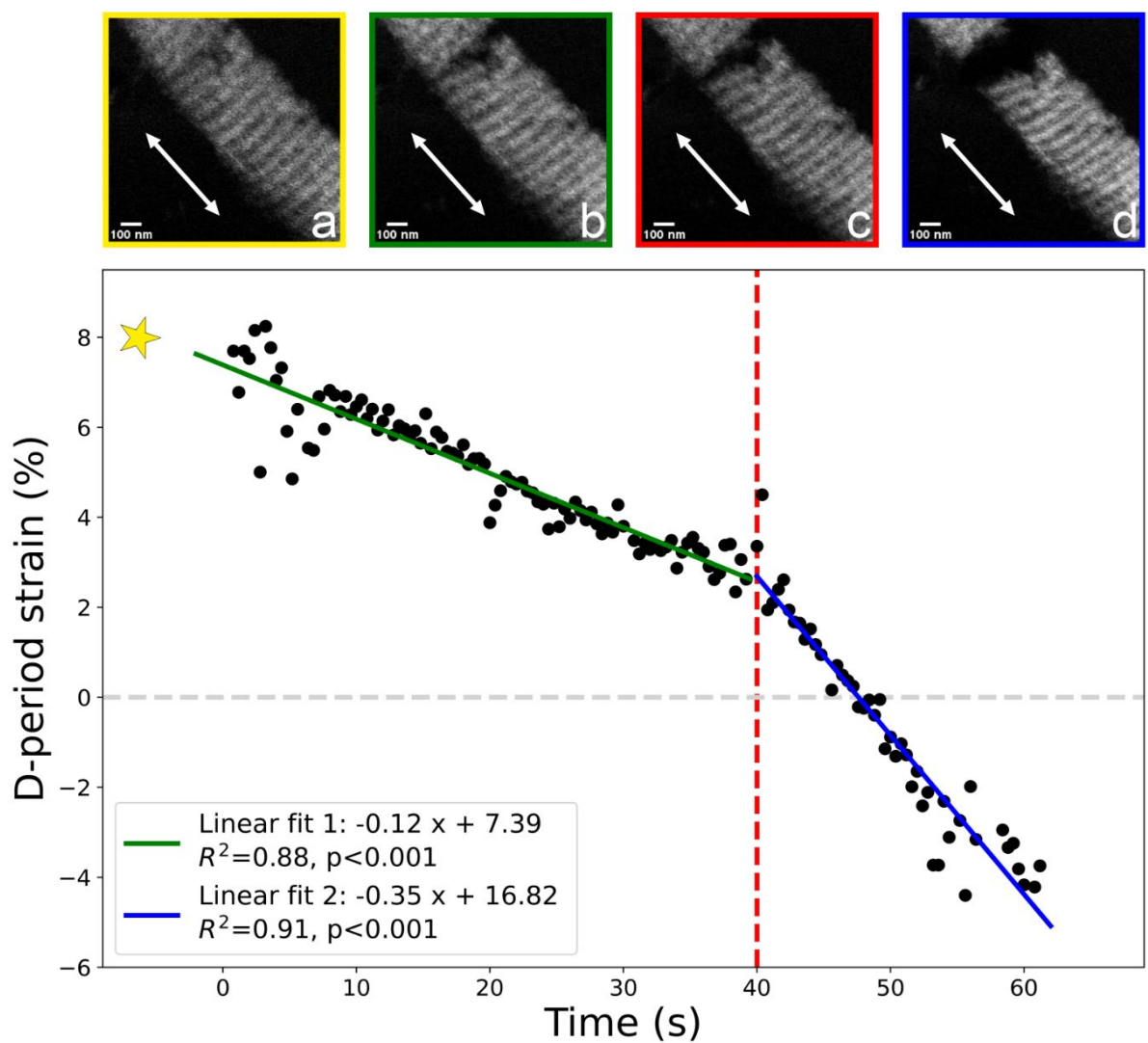

*Figure 6. Strain relaxation during crack propagation, along with the dark field TEM image sequences of the crack appearance (a-d). The D-period of the MCF prior to applied deformation was ≈69.5 nm and was taken as a zero-strain reference for the MCF D-period strain calculation. The star indicates the estimated moment of crack nucleation, which occurred prior to imaging. Initial crack appearance is visible in (a). The first phase of the strain relaxation aligns with the crack opening shown in (b) and continues up to the final MCF rupture at about 40 s (c). The second phase of the strain relaxation continues after the final MCF separation (d), reaching negative strain.*

# Discussion

In this work, a novel protocol for the MCFs extraction was developed, allowing multimodal TEM imaging of individual MCFs and subsequent *in situ* observations of their deformation mechanics. This was only possible following the newly established dropcasting protocol.

**Dropcasting** is an established and widely utilized technique for TEM sample preparation, wherein a small volume of sample suspension is deposited onto a TEM grid and allowed to dry. This method has successfully been employed to visualize the assembly of purified collagen molecules[16,31,32] and the formation of synthetic hydroxyapatite,[33,34] enabling high-resolution analysis of their morphology and organization.

To our knowledge, we are the first to apply dropcasting to native mineralized collagen fibrils (MCFs). While we initially attempted this procedure using bone samples, we encountered challenges as the MCFs tended to fragment into short pieces during the splitting process. However, when applying this method to mineralized turkey leg tendons, we successfully extracted longer samples exceeding 10 μm, while moderately removing non-collagenous proteins from the bulk material. In theory, this protocol could potentially be extended to less mineralized collagen fibrils, such as those found in antler bone, as well as non-mineralized collagen fibrils extracted from soft tissues, including tendons and muscles, following a thorough purification process.

From our observations, the average **D-period** of mineralized collagen fibrils was $68.64 \pm 0.16$ nm, slightly above the canonical ~67 nm value reported by Hodge and Petruska (1963).[5] Although 67 nm is widely cited as a defining structural feature of type I collagen, several studies have shown that D-spacing varies among fibril bundles.[35,36] This variability has been linked to native structural heterogeneity and to pathological changes, including estrogen-depletion-related osteoporosis.[37] In our samples, the reduction in D-period with higher Ca/P ratios may reflect both this inherent variability and a compaction effect from mineral infiltration. As mineral platelets occupy intrafibrillar gap zones and extend into overlap regions, they can restrict axial packing of collagen molecules, thus inducing axial contraction of collagen fibrils.[38] The pre-straining of the MCFs due to the vacuum exposure might have further introduced detected alterations from the canonical 67 nm.

Moreover, as was recently highlighted in the work of Shah,[39] the estimated mineralization from the Ca/P ratio may only serve as a proxy for bone mineral maturity. Mainly due to the oversimplified chemistry within bone since both $Ca^{2+}$ and $PO_4^{3-}$ ions are, in fact, partially substituted with $Mg^{2+}$, $Na^+$ and $CO_3^{2-}$. The suggested alternative would be the ratio of Ca+Mg+Na/P+C, which was not possible to estimate from our measurements due to the interference from the lacy carbon background signal. Additionally, the carbon signal in the gap zones may be influenced by edge effects, further complicating the analysis.

It is important to note that mineralized collagen fibrils (MCFs), like any biological tissue, are inherently heterogeneous, thus the native variations in mineralization is expected and in line with our observations. In the present work, samples were freshly prepared for each transmission electron microscopy modality to ensure that imaging was executed under consistent conditions immediately following extraction from the bulk material. Consequently, EDX mapping, 4D-STEM scans, and *in situ* stretching were conducted on different MCFs sourced from the same piece of bulk MTLT.

In this work, we used **flow line mapping** to study the orientation of hydroxyapatite crystallites in bone fibers.[40] Due to varying fibril thickness and heterogeneous mineralization, not all local orientations could be resolved. Although most imaged MCFs showed minerals well-aligned with the fibrillar axis, some fibrils exhibited substantial variations in mineral orientation, indicating a strong influence from extrafibrillar mineral (***Figure S 4***). In bulk bone tissue, extrafibrillar mineral is generally co-aligned with intrafibrillar mineral and the fibril axis.[6] Yet in our imaging of isolated MCFs, extrafibrillar mineral is likely redeposited on the fibril surface during preparation and may not reflect the native in-bulk arrangement. While intra- and extrafibrillar mineral cannot be distinguished in our 2D projection datasets, elucidating their 3D spatial organization to separate these contributions could be achieved through electron or atom probe tomography, as has been successfully applied in lamellar bone imaging.[6,8,41] The collected 4D-STEM scans nonetheless capture the overall orientation trends within individual fibrils, and these results and analysis pipeline will be valuable for future studies aimed at understanding mineralization progression within and along single MCFs, particularly when combined with emerging liquid-TEM capabilities.[42,43]

One limitation of our flowline analysis is that it focused solely on the (002) reflection. While this allows us to study the alignment of the c-axis relative to the fiber direction, it does not uniquely determine the orientation of crystals in 3D, leaving open questions about the relative rotation between the crystallites along the fiber. There are other 4D-STEM-based techniques that could address this question, such as automated crystal orientation mapping (ACOM).[44–46] In ACOM, individual experimental diffraction patterns are automatically indexed based on a known structure file to determine the alignment of crystals in 3D. This approach was not suitable for the datasets we acquired, given the limited number of reflections in each pixel and the low signal-to-noise ratio in our data, due to the low electron flux demanded by our beam-sensitive biocrystals. Moreover, there is natural variation in the composition and structure of hydroxyapatite, and we do not have a sufficiently good reference structure file for our material. Nonetheless, it may be possible to perform ACOM analysis in the future. Performing these measurements at cryogenic conditions would slow down damage,[47] and in this regime, it may even be possible to incorporate scanning precession electron diffraction to probe reflections at higher scattering angles that would greatly improve our analysis.[44,48] Lastly, correlative X-ray or electron diffraction experiments could provide the necessary references for solving the structures of our crystals, thereby facilitating ACOM.

Last but not least, we report observations from a single mineralized collagen fibril (MCF) subjected to ***in situ*** **tensile testing**. Notably, this represents the first direct visualization of crack propagation in an isolated MCF. The crack initiated and propagated within the collagen overlap region, characterized by lower mineral content, in agreement with numerical models predicting stress concentration in this region under uniaxial loading.[49] The progression of the crack revealed an interplay between the organic and mineral phases of the fibril, with a deflected crack path. This tropocollagen–hydroxyapatite interplay aids the energy dissipation and fracture resistance, similar to the well-documented bone toughening mechanisms at the macro-, meso-, and micro-scales.[50–53] By focusing on the deformation of individual fibrils in the TEM, we can begin to resolve the mechanical behavior within and between the organic and inorganic phases, providing new insights into the fundamental mechanisms of bone toughness.

D-period tracking enabled the estimation of local strain along the fibril near the crack. The observed linear strain relaxation was concurrent with the crack opening and ultimately resulted in negative strain values. These negative values are a direct consequence of our definition of zero D-period strain. Here, we used the D-period of the MCF, as measured prior to the applied deformation. However, the MCF appeared to be in a pre-strained state. Consequently, as the fibril relaxed toward its resting state, its D-period shortened below this initial reference, registering as a negative strain. Although strains recorded during crack progression reached at least 8.2%, the total strain alteration was even greater, amounting to 12.6% when considering the entire process down to the final fibril detachment. However, the true delta strain was likely even higher, since our measurements did not capture the precise moment of crack initiation when the fibril strain would have been at its peak. Our observations thus exceed existing molecular dynamics simulations, which suggest that mineralized fibrils may achieve a tensile strain of approximately 6.7%.[54]

Unlike nonmineralized fibrils, MCFs are challenging to extract in an intact state due to their native tight packing within the mineralized matrix. In an inspirational study of Hang and Barber on an antler bone, individual fibrils protruding from a fractured surface were mechanically pulled using the AFM cantilever, providing the stress-strain curves, but without direct visualization of the underlying deformation mechanisms.[55] At a higher structural level, MCF bundles from model tissue (MTLT) have been tested in compression in combination with *in situ* synchrotron imaging, revealing load-sharing mechanisms among the constituent fibers.[26] Beyond these examples, no prior work has demonstrated *in situ* mechanical testing of individual MCFs, largely because such experiments are considered technically impractical.

In the microscale bone tissue mechanics, recent advances have enabled testing of micron-sized specimens in uniaxial tension,[56,57] compression,[21,58-63] and other complex deformation fields.[64] However, the intricate sample preparation required for these approaches makes their application to isolated MCFs highly improbable. In contrast, numerous studies have investigated the mechanics of nonmineralized collagen fibrils using atomic force microscopy (AFM) or micro-electromechanical systems (MEMS) under various loading modes (nanoindentation, bending, tension)[65] and environmental conditions (air-dried, humidified, or submerged in aqueous solutions of varying ionic strength and pH).[65] Potentially, these testing approaches could be extended to individual MCFs by using the MTLT-based sample preparation strategy proposed in the present study

The *in situ* measurements presented here are inherently subject to beam-induced damage. In our organic samples with polymeric support films, the primary damage mechanism is radiolysis, while the effects of heating can be considered negligible. At an acceleration voltage of 300 keV, we observed that upon activating the electron beam, the support film underwent homogeneous expansion, leading to pre-stretching of the MCFs and potentially gradual disruption of fibrillar adhesion to the support film. This homogeneous stretching persisted for approximately 10 seconds; thereafter, the stretching of the film ceased. However, fibril sliding against the support film resulted in relaxation compared to the initial pre-strained state. This relaxation was confirmed by tracking the D-period of the MCF during static imaging (***Figure S 5***). Notably, the D-period relaxation observed during static imaging was an order of magnitude lower than the relaxation detected during the application of quasi-static stretching to the MCF: 10 nm/s and 1-2 nm/s, respectively. While this effect represents a non-negligible influence on MCF behavior, the reported deformation behavior is presented without any adjustments to avoid introducing potential bias.

The *in situ* mechanical testing approach for individual MCFs within an electron microscope presents significant potential. Following our preliminary experiments, we are optimistic regarding the future directions this technique could pursue. As discussed above, one notable limitation of our testing methodology is that fibrillar stretching is influenced by adhesion to the supporting film, which complicates any precise estimates of local forces. For future investigations, it may be advantageous to explore alternative designs for the tensile holder. For instance, a copper tensile holder with multiple smaller milled windows, as opposed to a single large central window, could be implemented. Utilizing 2x2 μm windows would eliminate the necessity for an underlying support film, while the ends of the fibrils can be fixed directly to the tensile grid using focused ion beam deposition. Additionally, another potential strategy for conducting *in situ* tensile tests could involve a push-to-pull device, aiding the real-time recording of forces during the experiment. However, our attempts to employ this method indicated difficulties in dropcasting the fibrils onto the designated region without contaminating the device.

In summary, this study employs mineralized turkey leg tendon (MTLT) as an excellent source for extracting mineralized collagen fibrils (MCFs) that closely resemble those found in bone. The dropcasting technique facilitates the successful transfer of individual MCFs onto TEM grids, allowing for detailed visualization of their native structural organization and composition through energy-dispersive X-ray spectroscopy (EDX) and 4D-STEM imaging. Furthermore, our pioneering tensile tests provide compelling evidence that bone toughening mechanisms initiate at the nanoscale, with observable crack deflection occurring at the level of individual MCFs, and remarkable tensile strains of up to 8%. These findings pave the way for future *in situ* mechanical testing using TEM, offering valuable insights into the mechanical properties of MCFs and potentially extending to other biological and architectured materials.

## Experimental Section

### Sample preparation

MTLT samples were prepared from a turkey leg, obtained from a local abattoir. Highly mineralized parts were dissected from the tendon bundles, mechanically cleaned from any soft tissue and further cut with a surgical scalpel. The resulting tendon pieces of about 1.5 mm in diameter and 4.0 mm in length were rinsed with PBS solution, dab dried and stored in the freezer at -20 °C. Before continuing sample manipulations, MTLT pieces were thrown in PBS solution (7.4 pH) for 24 hours, split along the main tendon axis via scalpel and ultrasonicated for 3 minutes in an Eppendorf with PBS. After that, the MTLT samples were transferred to Eppendorf tubes containing DI water to remove other non-collagenous components from the fibrils. The supernatant of the aqueous sample was examined under an optical microscope after dropcasting onto a heated Si wafer (~30 °C). If the supernatant quality was insufficient, the sample was transferred to a new Eppendorf tube with DI water, and the process was repeated until the supernatant contained a sufficient concentration of collagen fibers and a minimal amount of impurities. The resulting solution with mineralized collaged fibrils (MCF) was dropcasted on (i) a copper TEM grid with lacey carbon films for imaging or (ii) a copper tensile stripe with nitrocellulose support film in the milled central window (***Figure 2***).

## TEM modalities

Electron microscopes within the National Center for Electron Microscopy (NCEM) at Lawrence Berkeley National Laboratory were used in this study. STEM imaging and EDX mapping were carried out on an aberration-corrected FEI ThemIS microscope at 80 kV equipped with a Gatan direct electron K2 camera and a Bruker SuperX EDS system, providing quantitative EDS maps with a nanometer resolution.

4D-STEM imaging was carried out at TEAM I, the double-aberration-corrected FEI Titan microscope at 300kV with the Dectris Arina detector[66] operating in full frame mode. Gold nanoparticles were used to measure the elliptical distortion and to calibrate the reciprocal-space pixel size.

### *In situ* tensile tests

*In situ* tensile tests were performed on a TEAM I microscope with a Gatan single-tilt straining holder Model 654. The holder can elongate at a maximum rate of 2.0 µm s$^{-1}$ with a nominal drift rate of 1.5 nm mm$^{-1}$. The holder is purely displacement-controlled and does not measure force.

Custom-made tensile stripes made of 0.5 mm thin copper foil with a milled hole in the center were used. The nitrocellulose support film was formed on the striped window by 2% collodion in amyl acetate mixture. The MCF in solution were dropcasted on the support film and imaged *in situ* during tensile tests.

## Image processing and quantification

**EDX maps** were analyzed in Velox v.3.17. The intensity profile for C, N, O, P, Ca elements were corrected for the absorption and the k-factor and exported from each map to be analyzed in Python v.3.12. Intensity profiles were fitted with a linear combination of sinusoidal and a linear function (scipy curve_fit):

$$f(x) = a + bx + c\sin(dx + e) \qquad (1)$$

Where $a$ denotes the mean net intensity and $d$ its periodicity, matching with the physical D-period. The final D-period of each fibril is estimated as the mean $d$ value from all elements. The Ca/P ratio is estimated as the ratio of Ca to P mean intensities. The summary line plots with the fits for each fibril are shown in *Figure S 1*.

**4D-STEM analysis** was carried out in Python v3.12 using the open-source py4DSTEM v0.14.17 software package.[67] The analysis involved performing elliptical correction and calibration, followed by polar transformation. The angular position of the (002) diffraction spot was measured at each scan pixel in real space, with the resulting orientation vectors plotted as continuous lines on the real-space map, thereby illustrating the in-plane orientation of the crystallites. Their orientation is then compared to the in-plane orientation of the fibril, extracted through the intensity gradient. A schematic summary of the 4D-STEM data processing is provided in *Figure S 3*.

**The D-period tracking of the MCF during *in situ* stretching** was performed in Python, following the intensity profile fitting with eq. (1) as described for the EDX dataset. The initial, zero-strain D-period was determined from a static imaging series (30 frames) of the region of interest acquired prior to deformation. This reference value was then used to calculate the strain within the MCF during crack propagation.

## Acknowledgments

Authors acknowledge great support from the Molecular Foundry staff, especially Karen Bustillo, John Turner, and Shannon Ciston. Authors further thank Prof. Johann Michler for providing access to the EMPA Thun facility for preliminary work and Patrick Bühlmann for stimulating discussions. T.K. acknowledges substantial support from the "UniBE Short Travel Grants for (Post)Docs" from the University of Bern. Work at the Molecular Foundry was supported by the Office of Science, Office of Basic Energy Sciences, of the U.S. Department of Energy under Contract No. DE-AC02-05CH11231.

## Author Contributions Statement

T.K. conceived the project with guidance from P.S. and L.M.V. T.K., P.S. and L.M.V. performed the experiments. T.K. conducted the data analysis with the support of S.M.R., D.C., and P.S. T.K., A.M.M., and P.Z. acquired the funding. All authors contributed to the writing and review of the manuscript.

## Competing Interests

The authors declare that they have no competing interests.

## Data Availability

Available upon reasonable request.

# New avenues for characterizing individual mineralized collagen fibrils with transmission electron microscopy

Tatiana Kochetkova,[a*] Stephanie M. Ribet,[b] Lilian M. Vogl,[bde] Daniele Casari,[c] Rohan Dhall,[b] Philippe Zysset,[a] Andrew M. Minor,[bd] Peter Schweizer[be*]

[a]   ARTORG Center for Biomedical Engineering Research, University of Bern, Bern CH-3010, Switzerland;

[b]   National Center for Electron Microscopy (NCEM), The Molecular Foundry, Lawrence Berkeley National Laboratory, Berkeley, CA 94720, USA;

[c]   Laboratory for Mechanics of Materials & Nanostructures, Empa - Swiss Federal Laboratories for Materials Science and Technology, Thun CH-3603, Switzerland;

[d]   Department of Materials Science and Engineering, University of California Berkeley, CA 94720, USA;

[e]   Max Planck Institute for Sustainable Materials, Düsseldorf 40237, Germany

# 1. Summary of the EDX scans with the elemental intensity profile fitting and the output correlation of the D-period vs Ca/P ratio.

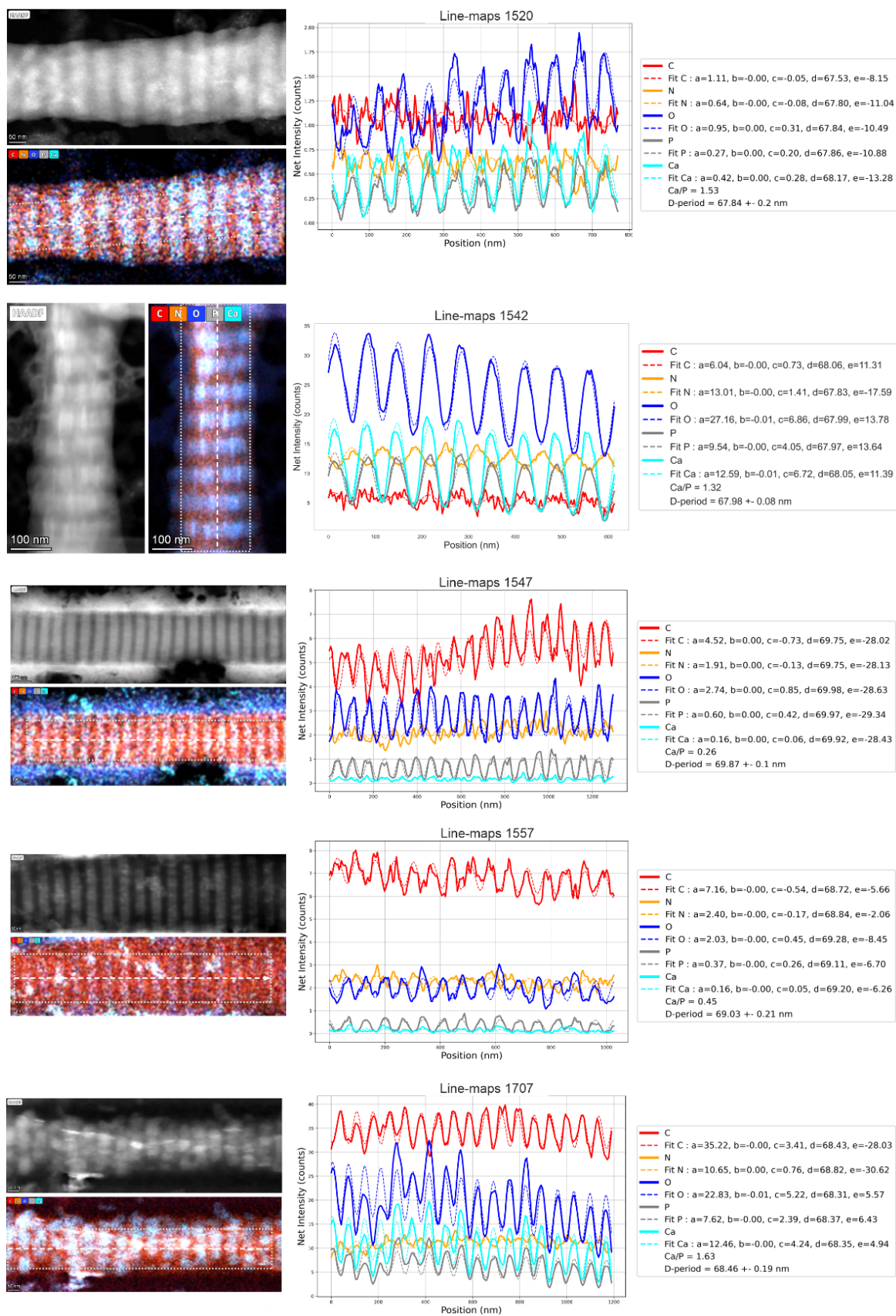

*Figure S 1. HAADF images and corresponding EDX scans for five fibers with the elemental intensity profile fitting.*

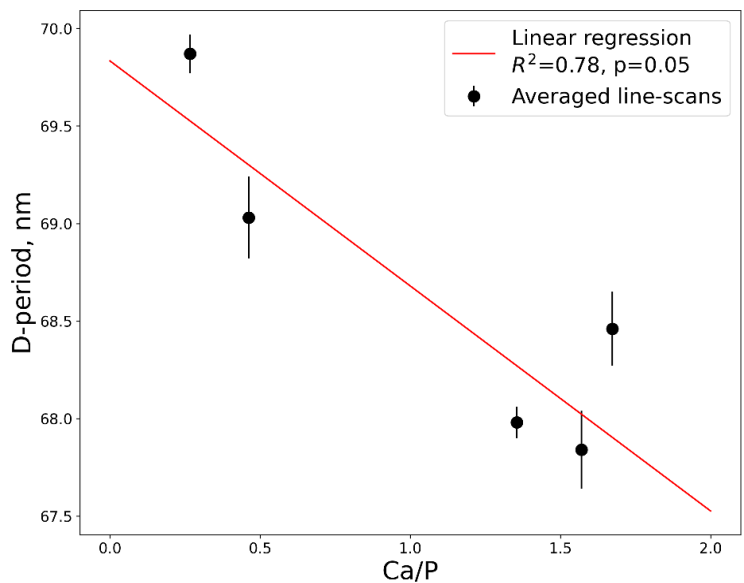

*Figure S 2. D-period negatively correlates with the Ca/P ratio.*

## 2. 4D-STEM data processing and summary

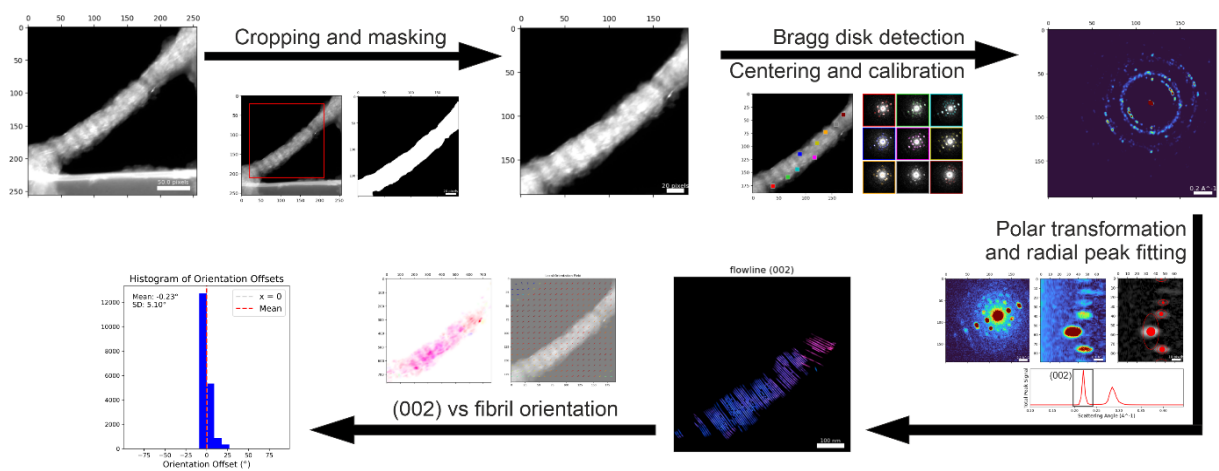

*Figure S 3. 4D-STEM data processing pipeline.*

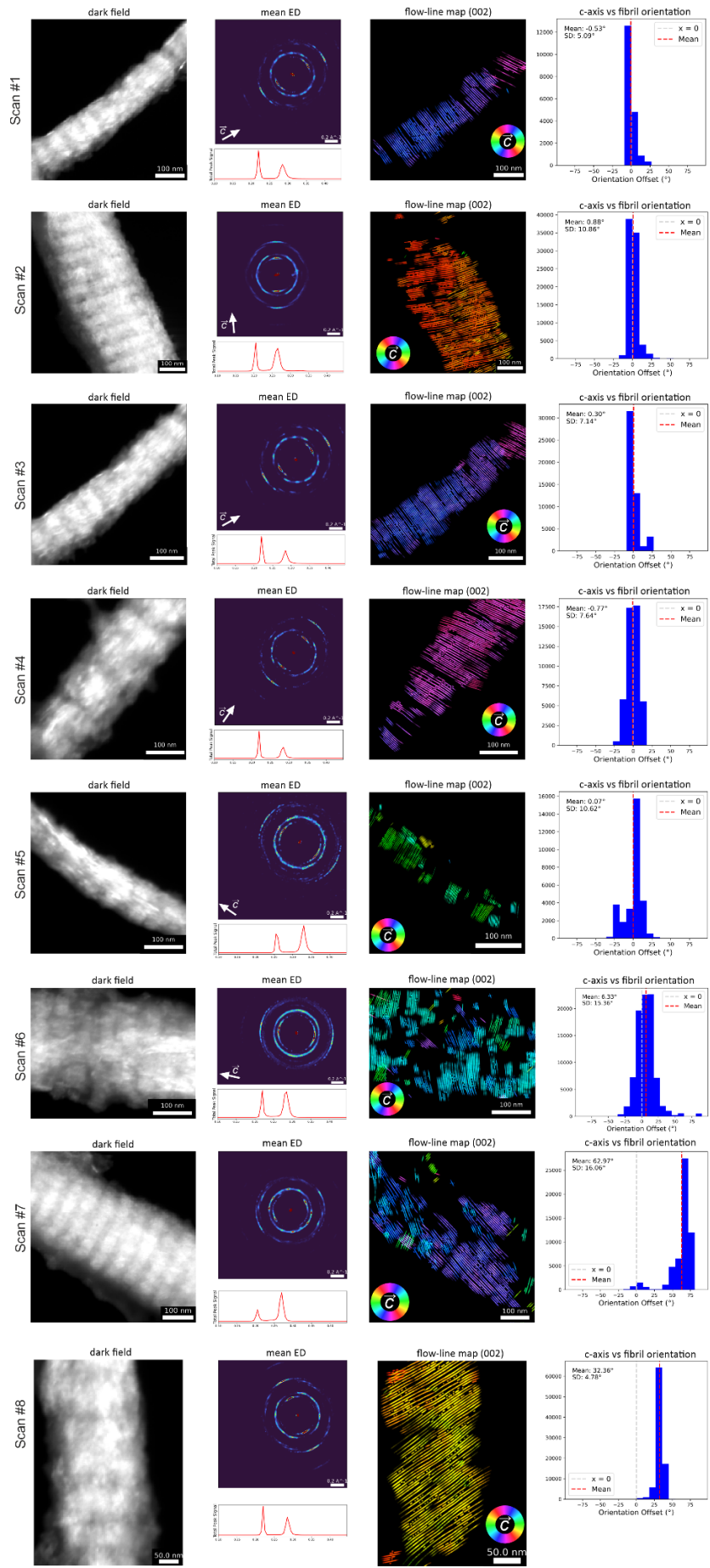

*Figure S 4. Summary of the flow line maps of all 4D-STEM scans, including (from left to right) reconstructed dark field image, mean electron diffraction and corresponding radial peaks, flow line map and resulting orientation variation between the fibril main axis and the c-axis of the mineral.*

## 3. *In situ* tensile tests

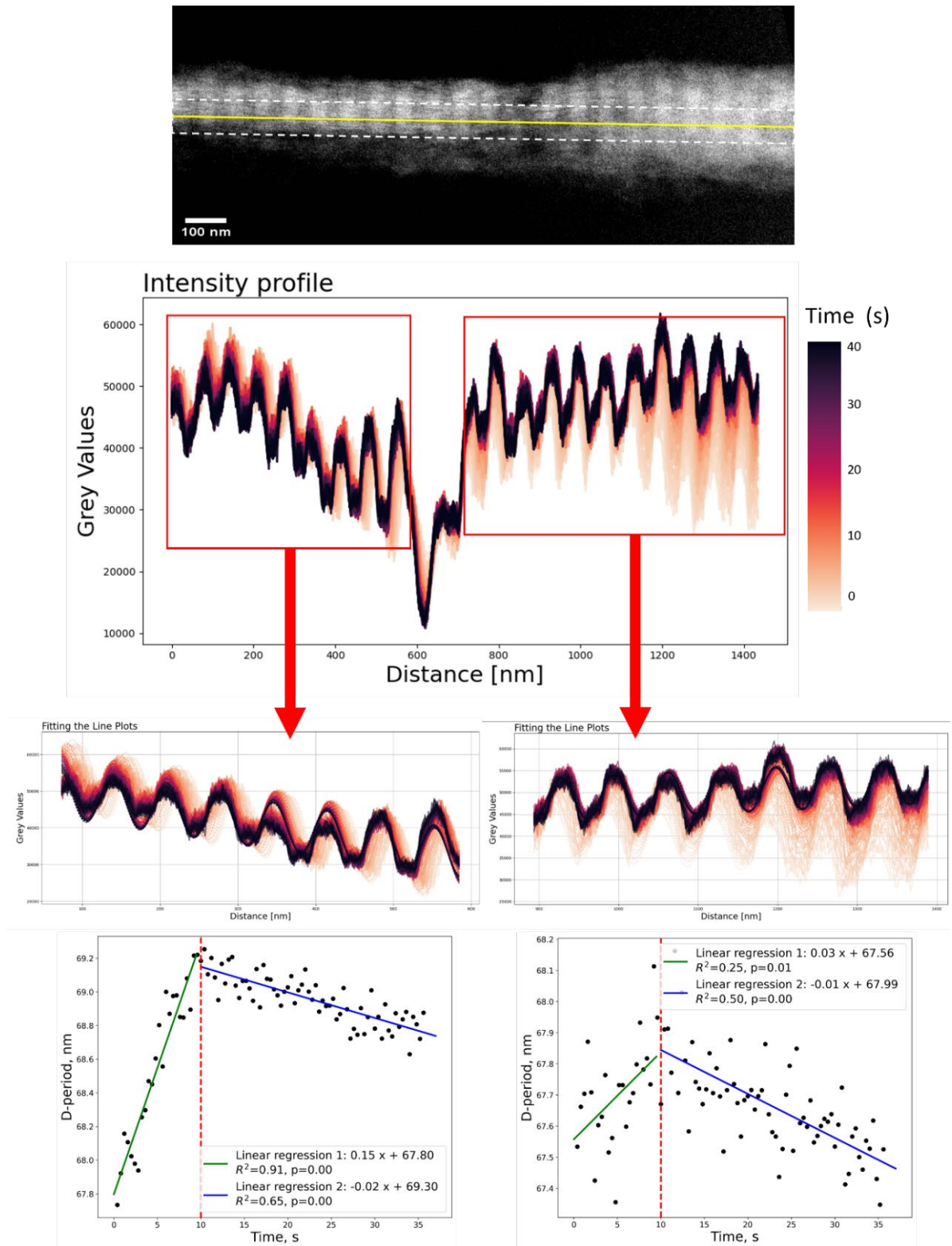

*Figure S 5. In situ TEM imaging of the MCF without external displacement from the straining holder. Note the two phases of the MCF's D-period evolution: (1) pre-stretching due to the homogeneous expansion of the support film upon electron beam activation (linear regression 1), followed by (2) relaxation, likely caused by the cessation of film expansion and the MCF's consequent sliding against the support film (linear regression 2).*